\documentclass[aip,jcp,reprint,longbibliography]{revtex4-1}
\usepackage{amsmath}
\usepackage{graphicx}

\newcommand{\dee}{\text{d}}

\newcommand{\Fatt}{F_{\text{att}}}
\newcommand{\fatt}{f_{\text{att}}}

\newcommand{\DG}{\Delta G}

\begin{document}

\title{A simple analytical formula for the free energy of ligand--receptor-mediated interactions}
\author{Stefano Angioletti-Uberti}
\author{Patrick Varilly}
\author{Bortolo M. Mognetti}
\affiliation{Department of Chemistry, University of Cambridge, Lensfield
  Road, CB2 1EW Cambridge, UK}

\author{Alexei V. Tkachenko}
\affiliation{Center for Functional Nanomaterials, Brookhaven National
  Laboratory, Upton, New York 11973, USA}

\author{Daan Frenkel}
\affiliation{Department of Chemistry, University of Cambridge, Lensfield
  Road, CB2 1EW Cambridge, UK}

\begin{abstract}
Recently~\cite{VarillyEtAl:2012}, we presented a general theory for
calculating the strength and properties of colloidal interactions mediated
by ligand-receptor bonds (such as those that bind DNA-coated colloids).  In
this communication, we derive a surprisingly simple analytical form 
for the interaction free energy, which was previously obtainable only via 
a costly numerical thermodynamic integration.  
As a result, the computational effort to obtain potentials of
interaction is significantly reduced.  Moreover, we can gain insight from
this analytic expression for the free energy in limiting cases.  In
particular, the connection of our general theory to other previous
specialised approaches is now made transparent.
This important simplification will significantly broaden the scope of our theory.
\end{abstract}

\maketitle

We consider a general system of many linkers, such as a solution of colloids coated with DNA strands that are capped
with reactive sticky-ends. At any given time, each linker~$i$
can bind at most one other distinct linker~$j$, with a free energy
change~$\DG_{ij}$ that depends on the polymer statistics of the linkers
(e.g., length, flexibility and grafting position).  In many cases, including
those of experimental relevance, the probability that linker~$i$ is unbound is
approximately independent of whether or not any other linker is also unbound.
Here, we show that in this limit, the free-energy of interaction of the
system is given by
\begin{equation}
\boxed{\beta\Fatt = \sum_i \ln p_i + \sum_{i<j} p_{ij}}
\label{eqn:FattExact}
\end{equation}
where $p_i$ is the probability that linker~$i$ is unbound and $p_{ij}$ is
the probability that linkers $i$~and~$j$ form a
bond. Previously\cite{VarillyEtAl:2012}, we showed that these quantities are
given by the unique physical solution to the following set of
self-consistent equations:
\begin{align}
p_{ij} &= p_i p_j e^{-\beta\DG_{ij}},\label{eqn:pij}\\
p_i &= 1 - \sum_j p_{ij}.\label{eqn:pi}
\end{align}

\begin{figure}
\begin{center}
\includegraphics[width=10.7cm
]{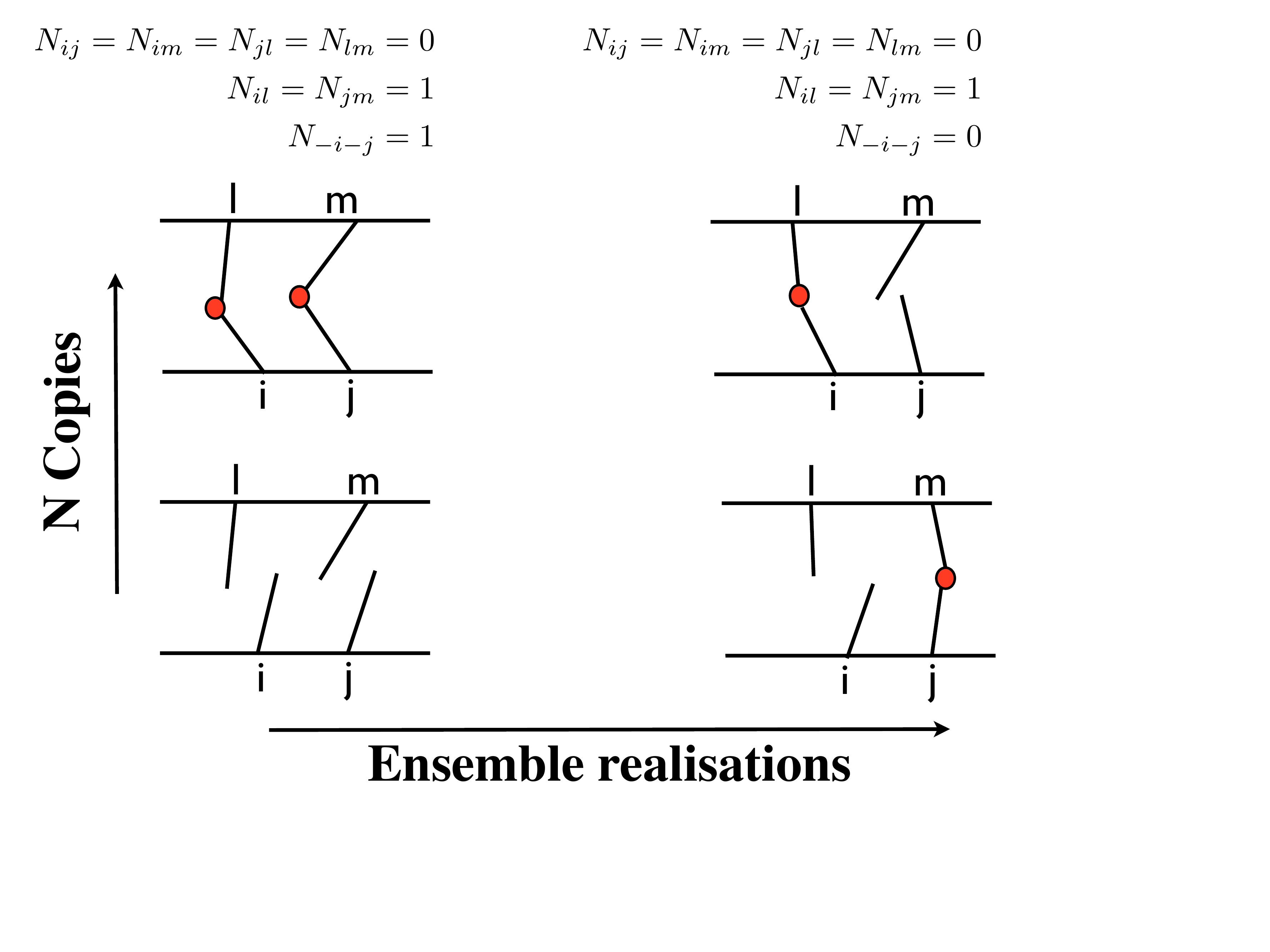}
\end{center}
\vspace{-1.5cm}
\caption{\label{fig:System} Two different realisations of an
  ensemble of two copies. Linkers are depicted as straight lines, and bonds are shown as filled circles.
  Although the numbers of bonds formed in each ensemble, $\{N_{ij}\}$, are equal, the number 
  of copies where both $i$ and $j$ are unbound, $N_{-i-j}$, differs.}
\end{figure}

In what follows, we first motivate the free energy expression in
Eq.~\eqref{eqn:FattExact} through a calculation that closely resembles that
for mixing entropy of solutions and gases.  This free energy is minimised
for the values of $\{p_i\}$~and~$\{p_{ij}\}$ that solve the self-consistent
conditions in Eqs. \eqref{eqn:pij}~and~\eqref{eqn:pi}. We then show that the
free energy in Eq.~\eqref{eqn:FattExact} is identical to that obtained
through the costly and numerical thermodynamic integration previously proposed. 
In the discussion section, we compare the performance of Eq.~\eqref{eqn:FattExact} with that of the
thermodynamic integration, establish the explicit connection between our Eq.~\eqref{eqn:FattExact} 
and a previous treatment of DNA-mediated colloid interactions\cite{DreyfusEtAl:2010}, and
state the analogous result to Eq.~\eqref{eqn:FattExact} for the mean-field
system of plates discussed in Ref.~\onlinecite{VarillyEtAl:2012}.

\section{Derivation of the Main Result}

We consider here an ensemble of $N$~independent copies of the
real system. In each copy, a different set of bonds forms between the
linkers (see Figure~\ref{fig:System}).  Let $N_i$ be the number of copies
where linker~$i$ is unbound, and let $N_{ij}$ be the number of copies
where $i$~and~$j$ are bound to each other.  Conversely, let $N_{-i,-j}$ be
the number of copies where both $i$~and~$j$ are unbound.  These quantities
are not independent: each linker $i$ is either unbound or unbound, so
\begin{equation}
N_i + \sum_j N_{ij} = N.
\end{equation}
The fraction of copies where $i$ is unbound ($f_i$), where $i$~and~$j$ are
bound to each other ($f_{ij}$), or where they are both unbound ($f_{-i,-j}$)
follow immediately:
\begin{align}
f_i &= N_i / N,\\
f_{ij} &= N_{ij} / N,\\
f_{-i,-j} &= N_{-i,-j} / N.
\end{align}
Let $Z(\{N_{ij}\})$ be the partition function of an ensemble under the constraint that 
each pair of linkers $i$~and~$j$ is bound to each other in exactly $N_{ij}$ copies.
A closed-form expression for $Z(\{N_{ij}\})$ can be constructed
recursively, by adding each bond one by one.  
For a given set of $\{N_{ij}\}$, we need to work out how $Z(\{N_{ij}\})$ changes upon
adding one more $i$-$j$ bond, which we do as follows.  
We call the set of realisations of the ensemble with $\{N_{ij}\}$ bonds the \emph{old} ensemble, and that of realisations 
with one more $i$-$j$ bond, the \emph{new} ensemble.  
In the old ensemble, there are $N_{-i,-j}$ copies where an $i$-$j$ bond can be added. 
In doing this, all the realisations of the new ensemble are generated, but not uniquely. 
For example, given two realisations, one with an $i-j$ bond in copy X but not in Y or Z, and one with an $i-j$ bond in Y but not in X or Z, 
the same final realisation can be obtained by adding an $i-j$ bond to Y in the former and X in the latter.
Conversely, in the new ensemble, we can generate a realisation of the old ensemble in $N_{ij}+1$ ways by
removing one of the $i$-$j$ bonds.
For example, the old realisation with an i-j bond in copy X but not Y or Z can be obtained by deleting the $i-j$ bond from a 
new realisation with an $i-j$ bond in X and Y but not Z, or from one with an $i-j$ bond in X and Z but not Y.
Since the number of ways of going from the 
old to the new ensemble is equal to the number of ways of going from the new to the old
ensemble, it follows that

\begin{align}
&Z(\cdots,N_{ij},\cdots) N_{-i,-j} e^{-\beta\Delta G_{ij}} \notag\\
&\qquad= Z(\cdots,N_{ij}+1,\cdots) (N_{ij}+1).
\label{eqn:balance}
\end{align}
The value of $N_{-i,-j}$ depends not just on the values of $\{N_{ij}\}$ but
on the details of how those bonds are distributed between copies (see Figure~\ref{fig:System}).  To 
remove this complication, we approximate the probability of $j$ being unbound as independent of whether
or not $i$ is also bound. Hence,
\begin{equation}
N_{-i,-j} = N f_{-i,-j} \approx N f_i f_j = \frac{N_i N_j}{N}.
\label{eqn:NijApprox}
\end{equation}
Neatly, this approximation allows us to
treat $N_{-i,-j}$ as a function of only $\{N_{ij}\}$.  
From the discussion above, we obtain an expression for the increase in
$Z(\{N_{ij}\})$ upon adding one $i$-$j$ bond to the system:
\begin{align}
\frac{Z(\cdots,N_{ij}+1,\cdots)}{Z(\cdots,N_{ij},\cdots)}
&\approx \frac{ e^{-\beta\Delta G_{ij}} N_i N_j}{N (N_{ij} + 1)}\notag\\
= &\frac{e^{-\beta\Delta G_{ij}}(N - \sum_k N_{ik}) (N - \sum_k N_{jk})}{N (N_{ij} + 1)}.
\end{align}
This recursion relation, and the fact that $Z=1$ when no bonds form, allows
us to write an approximate closed-form expression for $Z(\{N_{ij}\})$,
namely
\begin{equation}
Z(\{N_{ij}\}) \approx \prod_i \frac{N!}{(N - \sum_k N_{ik})!} \cdot
  \frac{1}{N^{\sum_{i<j} N_{ij}}} \cdot \prod_{i<j} \frac{e^{-\beta N_{ij}\Delta G_{ij}}}{N_{ij}!}.
\label{eqn:alexei}
\end{equation}
Using Stirling's approximation and Eq.~\eqref{eqn:alexei}, the free energy
per copy $\beta \fatt^* = -(1/N) \ln Z(\{N_{ij}\})$
is then given by 
\begin{widetext}
\begin{equation}
\beta\fatt^*(\{f_{ij}\}) =  
\sum\limits_{i<j} f_{ij} \beta\DG_{ij}
+ \sum\limits_i \bigl(1 - \sum\limits_j f_{ij}\bigr) \ln\bigl(1 - \sum_k f_{ik}\bigr)
 + \sum\limits_{i<j} f_{ij} \ln f_{ij} + \sum\limits_{i<j} f_{ij}.
\label{eqn:FattStarExpanded}
\end{equation}
\end{widetext}
Treating $\{f_{ij}\}$ as
continuous in the range $[0,1]$, the overall free energy per copy of the
ensemble, $\Fatt^*$, follows from a saddle-point approximation:
\begin{multline}
\beta\Fatt^* \equiv -\frac{1}{N}\ln\left[\int \Biggl(\prod_{i<j} N \dee f_{ij}\Biggr)\,
e^{-N\beta\fatt^*(\{f_{ij}\})}\right]\\
 =
\beta\fatt^*(\{\overline{f_{ij}}\}) + \mathcal{O}(\ln N / N) \approx
\beta\fatt^*(\{\overline{f_{ij}}\}),
\label{eqn:saddle}
\end{multline}
where the integration is over all positive values of $\{f_{ij}\}$ satisfying
$\sum_j f_{ij} \leq 1$ for all~$i$ and the values $\{\overline{f_{ij}}\}$ are
obtained by minimising the free energy per copy,
\begin{equation}
\frac{\partial\beta\fatt^*}{\partial f_{ij}}\biggr|_{\{\overline{f_{ij}}\}}
= 0.
\label{eqn:fijbar_defn}
\end{equation}
When $N\to\infty$, the values $\{\overline{f_{ij}}\}$ are precisely the
average values of $\{f_{ij}\}$.  Eq.~\eqref{eqn:fijbar_defn} implies that
$\{\overline{f_{ij}}\}$~and~$\{p_{ij}\}$ obey identical equations
(Eq.~\eqref{eqn:pij}), and so are equal:
\begin{equation}
\overline{f_{ij}} = p_{ij}.
\label{eqn:fij_is_pij}
\end{equation}
\subsection{Connection to thermodynamic integration}

The free energy $\Fatt^*$, defined in Eq.~\eqref{eqn:saddle}, is equal to
the free energy of the real system to the extent that the approximation in
Eq.~\eqref{eqn:NijApprox} is valid.  Since this is the same approximation
that we used previously\cite{VarillyEtAl:2012} to calculate the free energy in terms of a thermodynamic integral, 
$\Fatt$, it is reasonable to suppose that $\Fatt^*$~and~$\Fatt$ are equal.  We now show this explicitly.

In our original paper, we calculated the exact attractive free energy for
the real system of linkers using thermodynamic integration. Specifically, we
replaced $\beta\DG_{ij}$ by $\beta\DG_{ij} + \lambda$, whereupon the
probabilities $\{p_i\}$~and~$\{p_{ij}\}$ become functions of~$\lambda$.  We
then integrated the appropriate free energy derivative over the range $0
\leq \lambda < \infty$, and obtained
\begin{equation}
\beta\Fatt = -\int_0^\infty \dee\lambda\,\frac{\dee\beta\Fatt}{\dee\lambda}
= -\int_0^\infty \dee\lambda\, \sum_{i<j} p_{ij}(\lambda).
\label{eqn:ThermoInt}
\end{equation}
The same replacement of $\beta\DG_{ij}$ with $\beta\DG_{ij} + \lambda$ can
be made in the ensemble of $N$~copies.  In that case, using
Eqs. \eqref{eqn:FattStarExpanded}~and~\eqref{eqn:saddle}, we find that
\begin{equation}
\frac{\dee\beta\Fatt^*}{\dee\lambda} =
\sum_{i<j}
\frac{\partial\beta\fatt^*}{\partial f_{ij}}\Bigr|_{\{p_{ij}\}}
\frac{\dee p_{ij}}{\dee\lambda}
+ \frac{\partial\beta\fatt^*}{\partial \lambda}\Bigr|_{\{p_{ij}\}}.
\end{equation}
The first term vanishes owing to
Eqs. \eqref{eqn:fijbar_defn}~and~\eqref{eqn:fij_is_pij}, and the second term
follows immediately from Eq.~\eqref{eqn:FattStarExpanded}.  We then have
\begin{equation}
\frac{\dee\beta\Fatt^*}{\dee\lambda} =
\sum_{i<j} p_{ij} = \frac{\dee\beta\Fatt}{\dee\lambda}.
\end{equation}
Moreover, both $\Fatt^*$~and~$\Fatt$ are zero when $\lambda$ is infinite, so
the two quantities are equal for all~$\lambda$.

The previous result, together with Eqs.~\eqref{eqn:FattStarExpanded}--\eqref{eqn:fij_is_pij}, yields the following
closed-form expression for $\Fatt$:
\begin{widetext}
\begin{equation}
\beta\Fatt = \sum_{i<j} p_{ij} \beta\DG_{ij}
+ \sum_i \bigl(1 - \sum_j p_{ij}\bigr) \ln\bigl(1 - \sum_k p_{ik}\bigr)
+ \sum_{i<j} p_{ij} \ln p_{ij} + \sum_{i<j} p_{ij}.
\label{eqn:final_Fatt}
\end{equation}
This expression is, in fact, equivalent to the much more compact
Eq.~\eqref{eqn:FattExact}.  Concretely,
\begin{subequations}
\begin{align}
\beta\Fatt
&= \sum_{i<j} p_{ij} \beta\DG_{ij}
+ \sum_i (1 - \sum_j p_{ij}) \ln p_i
+ \sum_{i<j} p_{ij} \ln(p_i p_j e^{-\beta\DG_{ij}}) + \sum_{i<j} p_{ij},\\
&= \sum_i (1 - \sum_j p_{ij}) \ln p_i
+ \frac12\sum_{i,j} p_{ij} \ln p_i + \frac12\sum_{i,j} p_{ij} \ln p_j + \sum_{i<j} p_{ij},\\
&= \sum_i (1 - \sum_j p_{ij}) \ln p_i
+ \sum_{i,j} p_{ij} \ln p_i + \sum_{i<j} p_{ij},\\
&= \sum_i \ln p_i + \sum_{i<j} p_{ij}.
\label{eq:compact}
\end{align}
\end{subequations}
\end{widetext}

\section{Discussion}

\begin{figure}
\begin{center}
\includegraphics[width=\columnwidth]{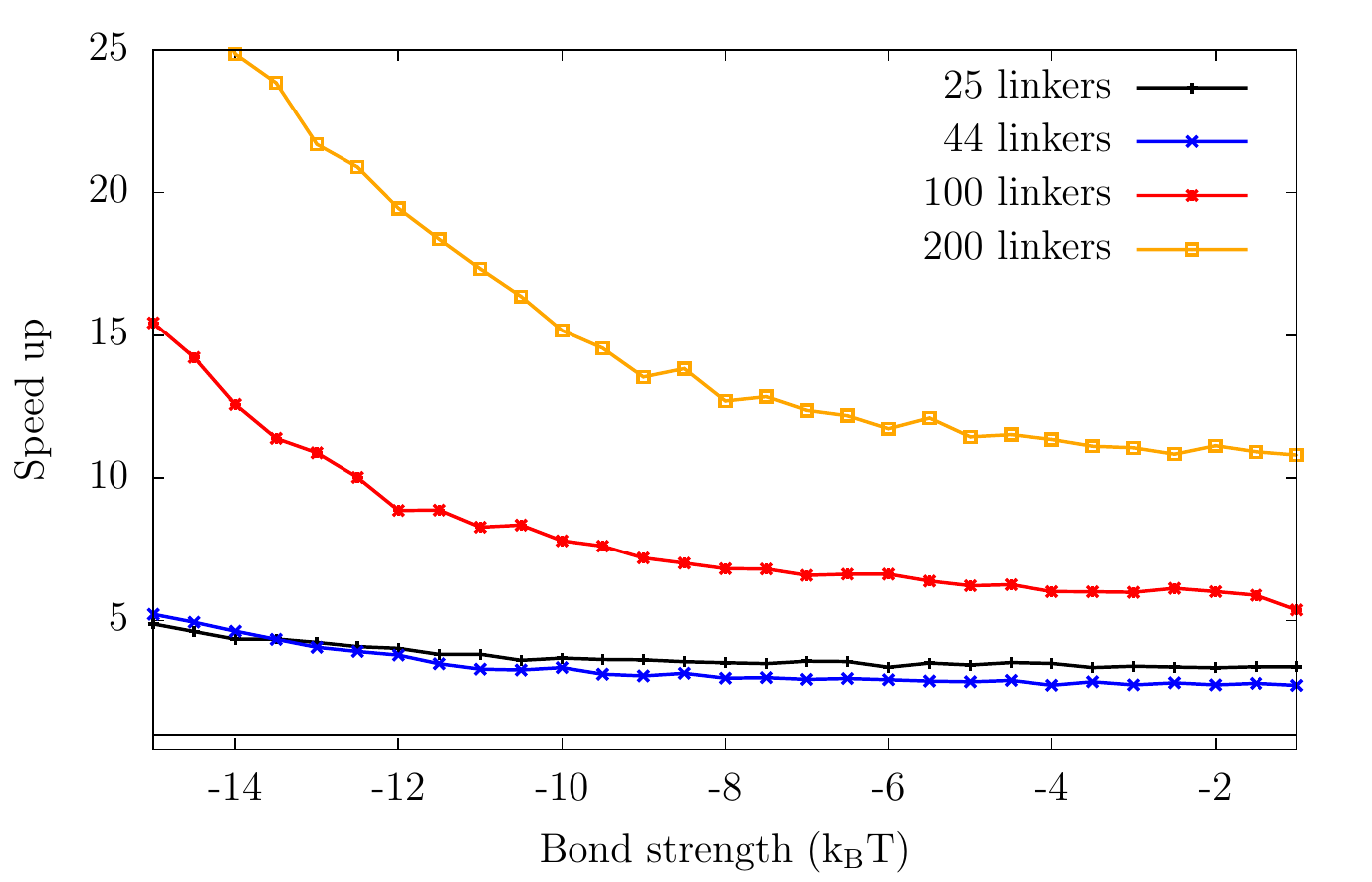}
\end{center}
\vspace{-0.5cm}
\caption{\label{fig:speedup}Computational speedups obtained
by using Eq.~\eqref{eqn:FattExact} vs. Eq.~\eqref{eqn:ThermoInt}, for a
system of parallel plates coated with complementary linkers.}
\end{figure}

Figure~\ref{fig:speedup} reports the typical computational speedups obtained
by using Eq.~\eqref{eqn:FattExact} versus our original thermodynamic
integral, Eq.~\eqref{eqn:ThermoInt}, for systems of $M$~linkers.  The
speedup is higher for larger~$M$ and for stronger bonds because each
evaluation of the thermodynamic integrand involves solving a system of
$M$~equations, and the size of the integration domain scales linearly with
the bond strength.  Typically, experimentally relevant regimes deal with
hundreds to tens of thousands of strands, a regime that can now be treated
exactly with Eq.~\eqref{eqn:FattExact}.

\begin{figure}
\begin{center}
\includegraphics[width=\columnwidth]{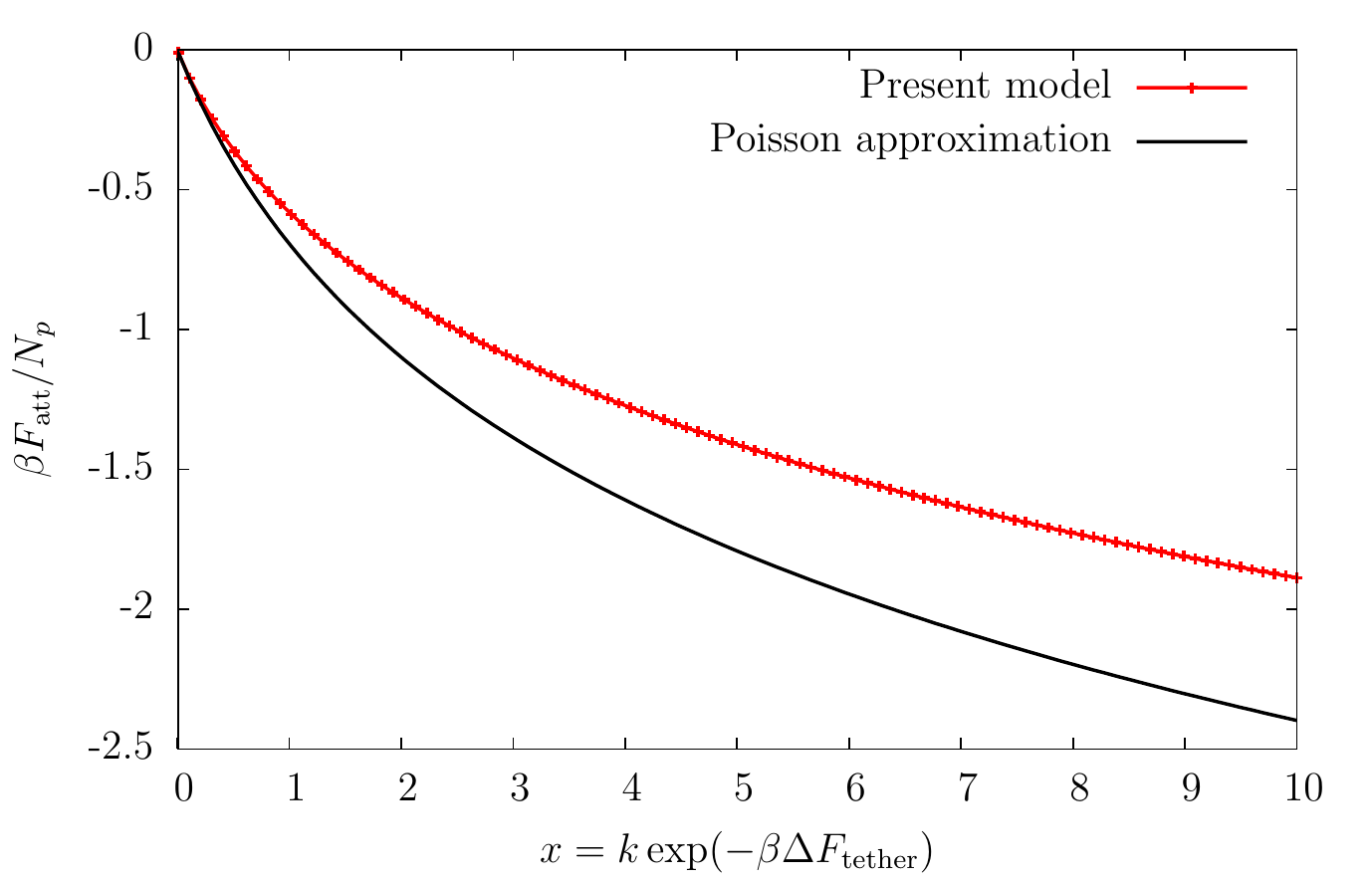}
\end{center}
\vspace{-0.5cm}
\caption{\label{fig:CompareModels} Free energy per linker in the ``Poisson
  approximation'' and the present model. Higher values of $x = k
  \exp(-\beta\Delta F_{\text{tether}})$ lead to higher bonding probabilities,
  either because bonds are stronger or because linkers have more binding
  partners. The two models agree in the ``weak binding regime'' ($x \ll
  1$), but disagree when correlations between neighbouring strands become
  significant.}
\end{figure}

In the limit of weak-bonds, where $p_{ij}$ is close to 0, 
we find that
\begin{align}
\beta\Fatt
&= \sum_{i} \ln p_i + \sum_{i<j} p_{ij} =\sum_{i} \ln (1-\sum_j p_{ij}) + \sum_{i<j} p_{ij} ,\notag\\
&\approx - \sum_{i,j} p_{ij} + \sum_{i<j} p_{ij} = -\sum_{i<j} p_{ij}.\notag
\label{eqn:OldModel}
\end{align}
This approximate result has been widely used by previous
authors\cite{BiancanielloKimCrocker:2005, LicataTkachenko:2006,
  DreyfusEtAl:2010, RogersCrocker:2011} under the name of the ``Poisson
approximation'' or the ``weak binding regime''.  However, in experiments
with micron-sized DNA-coated colloids, this approximation can be
significantly inaccurate\cite{MognettiEtAl:2012}.  At the nanoscale, where
high bond strengths are commonly used, the Poisson approximation is expected
to break down (see Fig.~\ref{fig:CompareModels} for comparison).
Eq.~\eqref{eqn:FattExact} is instead quantitatively accurate for bonds of
any strength\cite{VarillyEtAl:2012}.

Eqs. \eqref{eqn:FattExact}, \eqref{eqn:pij}~and~\eqref{eqn:pi} also make
explicit the connection between our theory and previous treatments.  For
example, Dreyfus et al.\cite{DreyfusEtAl:2010} model the attraction between
two DNA-coated spheres by first estimating the maximum number~$N_p$ of
linkers on each sphere that could form a bond with a linker on the second
sphere, and then assuming that each such linker can independently bind any
of $k$ linkers with an average free energy $\Delta F_{\text{tether}}$.  For
later convenience, we define a small expansion parameter~$x$ as
\begin{equation}
x \equiv k e^{-\beta\Delta F_{\text{tether}}},
\end{equation}
and note that the experiments in Ref.~\onlinecite{DreyfusEtAl:2010} take place
mostly in what the author call the ``weak-binding regime'', where $x \ll 1$
and $N_p x \gg 1$.  In their model, the partition function of the system is
\begin{equation}
Z \approx (1 + x)^{N_p},
\end{equation}
from which the free energy~$\Fatt$ follows,
\footnote{In
  Ref.~\onlinecite{DreyfusEtAl:2010}, the free energy of attraction 
  includes only states where at least one bond is formed, thus 
  $\beta\Fatt$ there is $-\ln[(1 + x)^{N_p} - 1]$. Our definition of
  $\Fatt$ is the free energy of the system of linkers, and includes the
  configuration where no bonds are formed}
\begin{align}
\beta\Fatt &\approx -\ln[(1 + x)^{N_p}] = -N_p\ln(1+x)\\
&\approx -N_p (x - x^2/2 + \cdots).
\label{eqn:OurModel}
\end{align}
In the present framework, which does not treat the linkers as binding
independently, every linker has the same probability~$p$ of being bound,
given by the solution to
\begin{equation}
p = \frac{1}{1 + x p} \Rightarrow p = \frac{\sqrt{1 + 4 x} - 1}{2x}.
\end{equation}

The free energy then follows from Eqs.
\eqref{eqn:FattExact},\eqref{eqn:pij}:
\begin{align}
\beta\Fatt &= N_p p^2 x + 2 N_p \ln p\\
&\approx -N_p (x - x^2 + \cdots)
\end{align}
Thus, our theory recovers the results of Dreyfus et al.~in the weak binding
regime.  However, there is significant disagreement already at second order
in~$x$, where linkers begin to compete for binding partners.

Finally, 
using the same procedure as in our original paper, we
can directly write an attractive free energy density 
for a pair of plates, treated at a more approximate, \emph{spatial} mean-field
level. In the notation 
of Ref.~\onlinecite{VarillyEtAl:2012}, 
\begin{equation}
\beta f_{\text{att}} = \frac12 \sum_{\alpha,\beta} \sigma_\alpha p_\alpha K_{\alpha\beta} p_\beta \sigma_\beta +
\sum_\alpha \sigma_\alpha \ln p_\alpha.
\end{equation}
This result also follows from a large-area limit of
Eq.~\eqref{eqn:final_Fatt} with random grafting points ~\cite{VarillyEtAl:2012,BortoloSoftMatter}.
We expect the simplification provided in this communication will boost
the use of our model for calculating interactions free-energy for general
ligand-receptor mediated systems.

\section{Acknowledgments}
This work was supported by the ERC 
Advanced Grant 227758, the Wolfson Merit Award 2007/R3 of the Royal Society of London
and the 
EPSRC Programme
Grant EP/I001352/1.  
P.V. has been supported by 
the 
Marie Curie International
Incoming Fellowship of the European Community's FP7 
PIIF-GA-2011-300045, and by a Tizard JRF
from Churchill College. Research carried out in part at the Center for Functional Nanomaterials, 
Brookhaven National Laboratory, which is supported by the U.S. Department of Energy, Office of Basic Energy Sciences, under Contract No. DE-AC02-98CH10886.

\end{document}